\documentclass{ecai}
\usepackage{times}
\usepackage{graphicx}
\usepackage{latexsym}
\usepackage{amsmath, amsthm, amssymb}
\usepackage{bm}
\usepackage[linesnumbered,ruled]{algorithm2e}
\usepackage{subfig}
\usepackage{multirow}
\usepackage[misc]{ifsym}

\begin{document}
\title{A Text-based Deep Reinforcement Learning Framework for Interactive Recommendation}

\author{Chaoyang Wang\institute{School of Computer Science and Technology, Huazhong University of Science and Technology, Wuhan, China, 430074. Email: \{sunwardtree,~georgeguo,~jianjunli,~panpeng,~guohuili\}@hust.edu.cn} \and Zhiqiang Guo$^{1}$ \and Jianjun Li$^{1,}$\thanks{Corresponding author.}  \and Peng Pan$^{1}$ \and Guohui Li$^{1}$}

\maketitle
\bibliographystyle{ecai}

\begin{abstract}
Due to its nature of learning from dynamic interactions and planning for long-run performance, reinforcement learning (RL) recently has received much attention in interactive recommender systems (IRSs). IRSs usually face the large discrete action space problem, which makes most of the existing RL-based recommendation methods inefficient. Moreover, data sparsity is another challenging problem that most IRSs are confronted with. While the textual information like reviews and descriptions is less sensitive to sparsity, existing RL-based recommendation methods either neglect or are not suitable for incorporating textual information. To address these two problems, in this paper, we propose  TDDPG-Rec, a Text-based Deep Deterministic Policy Gradient framework for interactive recommendation. Specifically, we leverage textual information to map items and users into a feature space, which greatly alleviates the sparsity problem. Moreover, we design an effective method to construct an action candidate set. By the policy vector dynamically learned from TDDPG-Rec that expresses the user's preference, we can select actions from the candidate set effectively. Through extensive experiments on three public datasets, we demonstrate that TDDPG-Rec achieves state-of-the-art performance over several baselines in a time-efficient manner.
\end{abstract}

\section{Introduction}

In the era of information explosion, recommender systems play a critical role in alleviating the information overload problem. Recently, interactive recommender system (IRS)~\cite{zhao2013interactive}, which continuously recommends items to individual users and receives their feedbacks to refine its recommendation policy, has received much attention and plays an important role in personalized services, such as Tik Tok, Pandora, and YouTube.

In the past few years, there have been some attempts to address the interactive recommendation problem by modeling the recommendation process as a multi-armed bandit (MAB) problem~\cite{li2010contextual,zhao2013interactive,wang2017factorization}, but these methods are not designed for long-term planning explicitly, which makes their performance unsatisfactory~\cite{chen2019large}. It is well recognized that reinforcement learning (RL) performs excellently in finding policies on interactive long-running tasks, such as playing computer games~\cite{mnih2015human} and solving simulated physics problems~\cite{lillicrap2015}. Therefore, it is natural to introduce RL to model the interactive recommendation process. In fact, recently there have been some works on applying RL to address the interactive recommendation problem~\cite{zheng2018drn,zhao2018deeppagewise,Zhao2018Recommendations,hu2018reinforcement}. However, most of the existing RL-based methods, including all Deep Q-learning Network (DQN) based methods~\cite{zheng2018drn,Zhao2018Recommendations,chen19generative} and most Deep Deterministic Policy Gradient (DDPG) based methods~\cite{hu2018reinforcement,zhao2018deeppagewise}, suffer from the problem of making a decision in linear time complexity with respect to the size of the action space, i.e., the number of available items, which makes them inefficient (or unscalable) when the action space size is large.

To improve efficiency, based on DDPG, Dulac-Arnold \emph{et al.}~\cite{dulac2015deep} proposed to first learn an action representation (vector) in a continuous hidden space, and then find the valid item by using a $k$ nearest neighbor search method. However, such a method ignores the importance of each dimension in the action vector. Moreover, it still needs to find the $k$ nearest-neighbors from the whole action space, which is time-consuming.
Recently, Chen \emph{et al.}~\cite{chen2019large} proposed a tree-structured policy gradient recommendation (TPGR) framework, within which a balanced hierarchical clustering tree is built over the items. Then, picking an item is formulated as seeking a path from the root to a certain leaf in the tree, which dramatically reduces the time complexity. But this method introduces the burden of building a clustering tree, especially when new items appear frequently, the tree needs to be reconstructed and this may cost a lot.

On the other hand, most exiting RL-based recommendation methods use the past interaction data, such as ratings, purchase logs, or viewing history, to model user preferences and item features~\cite{dulac2015deep,zhao2018deep,chen2019large}. A major limitation of such kind of methods is that they may suffer serious performance degradation when facing the data sparsity problem, which is very common in real-world recommendation systems. As well known, textual information like reviews by users and item descriptions provided by suppliers contains more knowledge than interaction data, and is less sensitive to data sparsity.  Nowadays, textual information has been readily available in many e-commerce and review websites, such as~\emph{Amazon} and \emph{Yelp}.  Thanks to the invention of word embedding, applying textual information for recommendation is possible, and there have been some successful attempts in conventional recommender systems~\cite{Zheng2017,Bauman2017,chin2018anr}. But for IRS, existing RL-based methods either neglect to leverage textual information, or are not suitable for incorporating textual information due to their unique structures for processing rating sequence.

In this paper, we propose a Text-based Deep Deterministic Policy Gradient framework for IRSs (TDDPG-Rec). Specifically, we utilize textual information and pre-trained word vectors~\cite{pennington2014glove} to embed items and users into a continuous feature space, which, to a great extent, alleviates the data sparsity problem. Then we classify users into several clusters by the K-means algorithm~\cite{agrawal1998automatic}.
Next, based on the thought of collaborative filtering, we construct an action candidate set, which consists of positive, negative and ordinary items that are selected based on the user's historical logs and classification results. Afterwards, we use a policy vector, which is dynamically learned from the actor part of TDDPG-Rec, to express the user's preference in the feature space. Finally, we use the policy vector to select items from the candidate set to form the action for recommendation.

\begin{figure}[t]
	\centering
	\includegraphics[width=0.80\columnwidth]{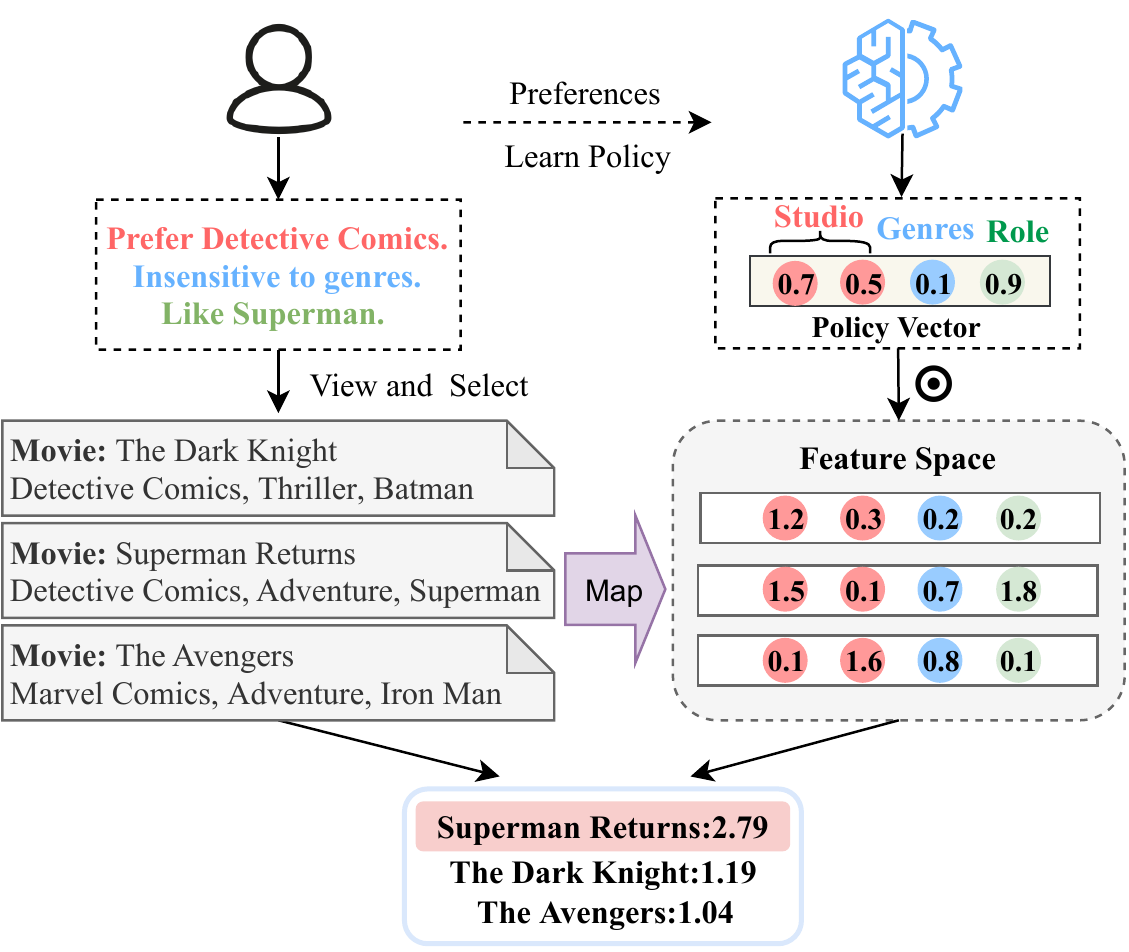}  
	\caption{An example for illustrating policy vector.}
	\label{fig1}
\end{figure}

Figure~\ref{fig1} gives an example for helping understand the policy vector. Suppose a user selects a movie according to the preference that can be represented as explicit policies such as {\it Prefer Detective Comics}, {\it Insensitive to genres} and {\it Like Superman}. By our method, a policy vector in the feature space, e.g., $(0.7, 0,5, 0.1, 0.9)$,  can be learned, where the value of each dimension represents how much emphasis the dimension is for this user. By conducting a dot product between the policy vector and the item vectors, we finally can choose the movie \emph{Superman Returns} with the highest score of $2.79$ for recommendation (assume Top-$1$ recommendation here).

Moreover, since it is too expensive to train and test our model in an online manner, we build an environment simulator to mimic online environments with principles derived from real-world data. Through extensive experiments on several real-world datasets with different settings, we demonstrate that TDDPG-Rec achieves high efficiency and remarkable performance improvement over several state-of-the-art baselines, especially for large-scale high-sparsity datasets. To sum up, our main contributions of this work are as follows:
\begin{itemize}
	\item  By utilizing textual information and pre-trained word vectors, we embed items and users into a continuous feature space to reduce the negative influence of rating sparsity.
	\item  We express the preferences of users by implicit policy vectors and propose a method based on DDPG to learn the policy vectors dynamically. Moreover, based on the thought of collaborative filtering, we classify users into several clusters and build the candidate set. The policy vector, combining with the candidate set, is used to select items that form an action, which reduces the scale of action space effectively.
	\item Extensive experiments are conducted on three benchmark datasets and the results verify the high efficiency and superior performance of TDDPG-Rec over state-of-the-art methods.
\end{itemize}

The remainder of this paper is organized as follows:  Section~\ref{sec: relatedwork} discusses related work; Section~\ref{sec: tddpg}  formally defines the research problem and details the proposed TDDPG-Rec model, as well as the corresponding learning algorithm; Section~\ref{sec: experiments} presents and analyzes the experimental results; Finally, Section~\ref{sec: conclusion}
concludes the paper with some remarks.

\section{Related Work}
\label{sec: relatedwork}

\subsection{RL-based Recommendation Methods}
RL-based recommendation methods usually formulate the recommendation procedure as a Markov Decision Process (MDP). They explicitly model the dynamic user's status and plan for long-run performance~\cite{tan2017neural,zheng2018drn,hu2018reinforcement,zhao2018deeppagewise,Zhao2018Recommendations,chen19generative,Zhao2019DRLSearch}. As mentioned earlier, most existing RL-based methods suffer from the large-scale discrete action space problem.

To address the large-scale discrete action space problem in IRS, there are some good attempts. Dulac-Arnold \emph{et al.}~\cite{dulac2015deep} proposed to leverage prior information about the actions to embed them in a continuous space to generate a proto-action, and then find a set of discrete actions closest to the proto-action as the candidate in logarithmic time via a $k$ nearest neighbor search. This method ignores the negative influences of the dimensions that users do not care about, which makes it fail to find proper actions sometimes. Moreover,  the $k$ nearest-neighbor search needs to be conducted on the whole action space, which still surfers a high runtime overhead. Zhao \emph{et al.}~\cite{zhao2018deep} used the actor-part of the Actor-Critic network to gain $k$ weight vectors, each of which can pick up a maximum-score item from the remaining items. But the relationship of these vectors is blurry, which causes the order of the $k$ items cannot be explained. Based on DPG, Chen \emph{et al.}~\cite{chen2019large} proposed a tree-structured policy gradient recommendation (TPGR) framework. In TPGR,  a balanced hierarchical clustering tree is built over all the items. Then, making a decision can be formulated as seeking a path from the root to a certain leaf in the clustering tree, which reduces the time complexity significantly. But this method can only support Top-$1$ recommendation. Moreover, when new items appear frequently,  the clustering tree needs to be reconstructed, which incurs extra cost.

\subsection{Textual Information for Recommendation}
Most of the recommendation models (including RL-based ones) that merely exploit interaction matrix usually face the data sparsity problem. A large amount of knowledge in the textual information can potentially alleviate the data sparsity problem~\cite{Zheng2017}. The development of deep learning in natural language processing makes it possible for using textual information that human beings can understand to enhance the recommendation performance~\cite{Zheng2017,Bauman2017,chin2018anr,Cheuque2019steam}.
Reviews and descriptions are the most important textual information in recommender systems. The reviews, which contain users' attitudes, and the descriptions, which contain items' advantages, along with the ratings, can show the preferences of users.
There are works that use sentiment analysis~\cite{Bauman2017}, convolutional neural networks~\cite{Zheng2017,deng2018neural} and pre-trained word vectors on large corpora~\cite{chin2018anr}, to get vectors from textual information. These vectors are then incorporated into the proposed model to improve recommendation performance.

IRS also suffers from the rating sparsity problem, but so far, we
are not aware of any recommendation method for IRS that utilizes
textual information. Most existing RL-based methods for IRS either neglect to incorporate with textual information, or have difficulty in utilizing textual information, since they use rating sequence, which has time-related structures, as the input of their model~\cite{zheng2018drn,chen2019large}. Note that in the domain of conversational recommender system (CRS), Basile~\emph{et al.}~\cite{basile2018deep} proposed a framework that combines deep learning and reinforcement learning and uses text-based features to provide relevant recommendations and produce meaningful dialogues. But different from CRS, in our RL-based method for IRS, the textual information is utilized to learn the implicit long-term preferences, not the proactive immediate needs of users.

\section{Proposed Method}
\label{sec: tddpg}

\subsection{Problem Formulation}

We consider an interactive recommendation system with $N$ users  $\{u_1, u_2, \ldots, u_N\}$  and $M$ items $\{v_1, v_2, \ldots, v_M\}$, and use $Y \in \mathbb{R}^{N \times M}$ to denote the rating matrix, where $y_{i,j}$ is the rating of user $u_i$ on item $v_j$. This kind of interactive Top-$k$ recommendation process can be modeled as a special Markov Decision Process (MDP), where the key components are defined as follows.

\begin{itemize}
	\item \textbf{State.}  Use $\mathrm{S}$ to denote the state space. A state $s \in \mathrm{S}$ is defined as the possible interaction between a user and the recommender system, which can be represented by $n_s$ item vectors.
	
	\item \textbf{Action.} Use $\mathrm{A}$ to denote the action space. An action $a \in \mathrm{A}$ contains $n_a$ ordered items, each of which is represented by a vector, for recommendation.
	\item \textbf{Reward function.} After receiving an action $a$ at state $s$, our environment simulator returns a reward $r$, which reflects the user's feedback to the recommended items. We use $\mathcal{R}(s, a)$ to denote the reward function.
	\item \textbf{Transition.}  In our model, since the state is a set of item vectors, once the action is determined and the user's feedback is given, the state transition is also determined.
\end{itemize}

Consider an agent that interacts with the environment $E$ in discrete timesteps. At each timestep $t$, the agent can receive a state $s_{t}$ by observing the current environment,  then it takes an action $a_{t}$ and gets a reward $r_{t}$.
An agent's behavior is defined by a policy $\pi$, which maps states to a probability distribution over the action, i.e., $\pi : \mathrm{S} \to \mathcal{P}(\mathrm{A})$.
Based on the above notations, we can define the instantiated MDP for our recommendation problem,  $\mathcal{M}=\langle \mathrm{S}, \mathrm{A}, \mathcal{R}, \mathcal{P}, T, \gamma\rangle$, where $T$ is the maximal decision step, and $\gamma$ is the discount factor. Our objective in this work is to learn a policy $\pi$ that maximizes the expected discounted cumulative reward.

\subsection{Framework Overview}
\begin{figure}[t]
	\centering
	\includegraphics[width=0.98\columnwidth]{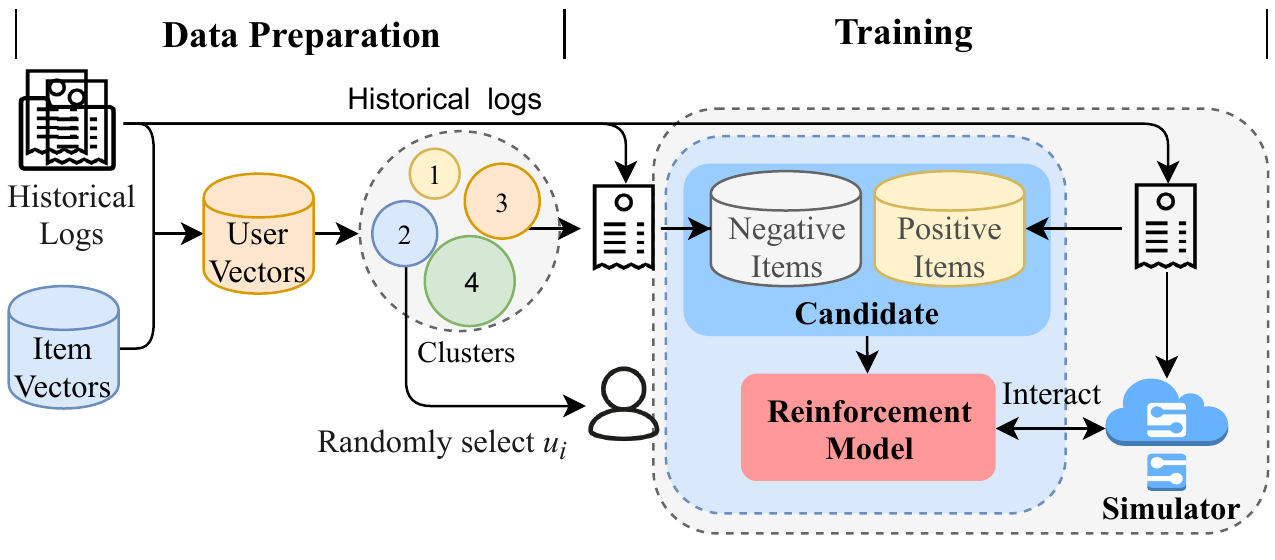}
	\caption{Framework Overview.}
	\label{fig: overview}
\end{figure}

Figure~\ref{fig: overview} gives an overview of our framework, which contains two major steps: data preparation and training. In data preparation, we first embed items to get item vectors by leveraging textual information.  Based on the derived item vectors and users' historical logs, we can embed users into the same feature space. Next, we classify the users into several clusters by K-means. In the training phase, we train a unique model for each cluster, with the objective of implementing a more personalized recommendation. Take cluster $2$ for an example, we randomly select a user $u_i$ from it.  Based on the historical log of $u_i$ and the user classification results, we sample positive, negative and ordinary items for $u_i$ to construct a candidate set, which later will be used in the reinforcement model for action selection. Our reinforcement model is based on DDPG, which interacts with the simulator that is based on historical logs to learn the inner relationship among all possible states and actions. The training phase will stop when the model loss reaches stable.

\subsection{Embedding with Textual Information}
Textual information like descriptions and reviews is important for decision making, we build vectors based on them.  Item vectors are calculated by the word vectors from GloVe.6B\footnote{http://nlp.stanford.edu/data/glove.6B.zip} (trained on  Wikipedia 2014 and Gigaword 5). Note that the descriptions and reviews contain many meaningless words, we remove them in advance by comparison with the Long Stopword List\footnote{https://www.ranks.nl/stopwords}.
Using $\textbf{\textrm{u}}_i$ and $\textbf{\textrm{v}}_j$ to denote the vectors of representing user ${u}_{i}$ and item $v_j$, respectively, the item vector $\textbf{\textrm{v}}_j$ can be  computed by,
\begin{equation}
\textbf{\textrm{v}}_j = \frac{1}{n_d}\sum\nolimits_{p=1}^{n_d} \textbf{\textrm{w}}^D_p + \frac{1}{n_r}\sum\nolimits_{q=1}^{n_r} \textbf{\textrm{w}}^R_q
\end{equation}
where $\textbf{\textrm{w}}^D_p$ and $\textbf{\textrm{w}}^R_q$ are the vectors of the words from descriptions and reviews, respectively, and $n_d$ and $n_r$ denote the corresponding numbers of them.
The word vectors with similar semantics have closer Euclidean distance than the word vectors with large semantics differences \cite{pennington2014glove}, which ensures that items with similar reviews and descriptions are closer to each other.

Given a user $u_i$ and one of its historical logs, if the corresponding rating is greater than a given bound $y_b$ (e.g., $y_b = 2$ in a rating system with the highest rating $5$), then the log is regarded as positive; Otherwise, it is negative. We use $\mathcal{V}^p_{u_i}$ and $\mathcal{V}^n_{u_i}$ to denote the set of items that are in $u_i$'s positive and negative historical logs, respectively. After obtaining all the item vectors, we can calculate user vector $\textbf{\textrm{u}}_i$ by normalizing the summation of the items' vectors that appear in $\mathcal{V}^p_{u_i}$, i.e.,
\begin{equation}
\textbf{\textrm{u}}_i = \frac{1}{n_v}\sum\nolimits_{l=1}^{n_v} \textbf{\textrm{v}}_l
\end{equation}
where $n_v$ denotes the number of items in $\mathcal{V}^p_{u_i}$. In this way, we embed users and items in a same feature space.

\subsection{Construction of the Candidate Set}
In Top-$k$ recommendation, the state is defined as a set of $n_s$ items. So there are a total of $A_{M-n_s}^k$ (note here $A_{M-n_s}^k$ is a permutation) actions that can be chosen as an action. With the increase of the number of items ($M$), the scale of the action space will increase rapidly. Based on the assumption that the preferences can be obtained by a set of items that users like and dislike, we pick up the positive and negative items to build a candidate set $c$. Additionally, to maintain generalization, we add some ordinary items in the candidate set.

For user $u_i$, we sample positive items from $\mathcal{V}^p_{u_i}$, negative items from $\mathcal{V}^n_{u_i}$, and ordinary items by random. Since users usually skip the items that they do not like, the negative items in $\mathcal{V}^n_{u_i}$ are rare~\cite{Marlin2009}. Based on the reverse thought of collaborative filtering, i.e., the more differences between two users, the more possible that the one's likes are another's dislikes, we classify users into several clusters by K-means~\cite{agrawal1998automatic} to supplement negative items.
Specifically, we denote the set of items that appear in the positive historical logs of users in cluster $l$ as $\mathcal{V}^p_{cl_l}$ (user $u_i$ belongs to cluster $cl_l$), and use ${cl_{l}^f}$ to denote the cluster that has the farthest distance from the current cluster $cl_l$. If the negative items in $\mathcal{V}^n_{u_i}$ are not enough, the rest negative items will be selected from $\mathcal{V}^p_{cl_{l}^f} - (\mathcal{V}^p_{cl_{l}^f} \cap (\mathcal{V}^p_{cl_{l}} \cup \mathcal{V}^n_{u_i}))$. In this way, we can reduce the scale of the action space from $M-n_s$ to $n_c$, where $n_c$ is the number of items in the candidate set $c$.

Algorithm 1 shows the detail of the construction for the candidate set, in which the positive items account for no more than $\alpha$ percent (line $1$), and the negative and ordinary items each share $50\%$ of the remaining part of $n_c$ (line $7$).
In the training phase, since constructing a candidate set only contains some simple operations, such as randomly select and merge, and the size of candidate set is always fixed, it is not difficult to see the time complexity of Algorithm 1 is constant.

\begin{algorithm}[t]
	\caption{Candidate set construction for ${u}_{i}$}
	\KwIn{$n_c$, $\alpha$, $\mathcal{V}^p_{u_i}$, $\mathcal{V}^n_{u_i}$, $\mathcal{V}^p_{cl_{l}}$ and $\mathcal{V}^p_{cl_{l}^f}$.}
	\KwOut{Candidate set $c$.}
	Initialize $c = \emptyset$, \ $n_{pos} = \lfloor n_c \times \alpha \rfloor$\;
	\eIf{$n_{pos} \; \leq \; |\mathcal{V}^p_{u_i}|$}
	{
		$c \leftarrow$ randomly select $n_{pos}$ items from $\mathcal{V}^p_{u_i}$\;
	}
	{
		$n_{pos} = |\mathcal{V}^p_{u_i}|$;~~~$c \leftarrow \mathcal{V}^p_{u_i}$\;
	}
	$n_{neg} = \lfloor{(n_c - n_{pos}) / 2}\rfloor$\;
	\eIf{$n_{neg} \; \leq \; |\mathcal{V}^n_{u_i}|$}
	{
		$c \leftarrow c \; \cup $ randomly select $n_{neg}$ items from $\mathcal{V}^n_{u_i}$\;
	}
	{
		$n_{neg} = n_{neg} - |\mathcal{V}^n_{u_i}|$;~~~ $c \leftarrow c \cup \mathcal{V}^n_{u_i}$\;
		$\mathcal{V}_{neg} \leftarrow \mathcal{V}^p_{cl_{l}^f} - (\mathcal{V}^p_{cl_{l}^f} \cap (\mathcal{V}^p_{cl_{l}} \cup \mathcal{V}^n_{u_i}))$\;
		\eIf{$n_{neg} \; \leq \; |\mathcal{V}_{neg}|$}
		{
			$c \leftarrow c \; \cup$  randomly select $n_{neg}$ items from $\mathcal{V}_{neg}$\;
		}
		{
			$c \leftarrow c \cup \mathcal{V}_{neg}$\;	
		}
	}
	$n_{ord} = n_c - |c|$\;
	$c \leftarrow c \; \cup$ randomly select $n_{ord}$ items not in $c$\;
	return $c$\;
\end{algorithm}

\subsection{Architecture of TDDPG-Rec}
The goal of a typical  reinforcement learning model is to learn a policy $\pi$ that can maximize the discounted future reward, i.e., the Q-value, which is usually estimated by the state-action value function $Q^{\pi}(s_{t}, a_{t})$.
Combined with deep neural networks, there are many algorithms that try to approximate $Q^{\pi}(s_{t}, a_{t})$. Among them, DDPG, a model-free, off-policy actor-critic algorithm, which combines the advantages of DQN \cite{mnih2015human} and DPG \cite{Silver2014}, can concurrently learn policy and $Q^{\pi}(s_{t}, a_{t})$ in high-dimensional, continuous action spaces by using neural network function approximation~\cite{lillicrap2015}. We use DDPG in our model, and Figure 3 shows the architecture of it.

In each timestep $t$, the actor network takes a state $s_t$ as input. By a multiple-layer perceptron (MLP) network, we can learn a continuous vector, which we term as the policy vector, denoted by $\textbf{\textrm{p}}_t$. The critic network takes state $s_t$  and policy vector $\textbf{\textrm{p}}_t$ as input. By an MLP, it can learn the current Q-value to evaluate $\textbf{\textrm{p}}_t$. As mentioned in Figure 1,  $\textbf{\textrm{p}}_t$ represents a user's preferences in the feature vector space, it is a continuous weight vector that can measure the importance of each dimension. Combining with the candidate set $c_t$, we can get $n_a$ items with the highest score, each of which is denoted by $Score(v_i)$ and,
\begin{equation}
Score({{v}_{i}})= \textbf{\textrm{p}}_t^T \textbf{\textrm{v}}_i
\end{equation}
Moreover, to cover the action space to a large extent, the candidate set is randomly generated at each time step.

Note that the actions in IRSs are discrete. In our method, when embedding the items, we have mapped the discrete actions into a continuous feature space, where each item is represented by a feature vector. Then, by conducting the dot product between $\textbf{\textrm{p}}_t$ and the item vectors in $c_t$, we can select the actions from a discrete space. In this way,  our method can overcome the gap between discrete actions in IRSs and continuous actions in DDPG.

\begin{figure}[t]
	\centering
	\includegraphics[width=0.95\columnwidth]{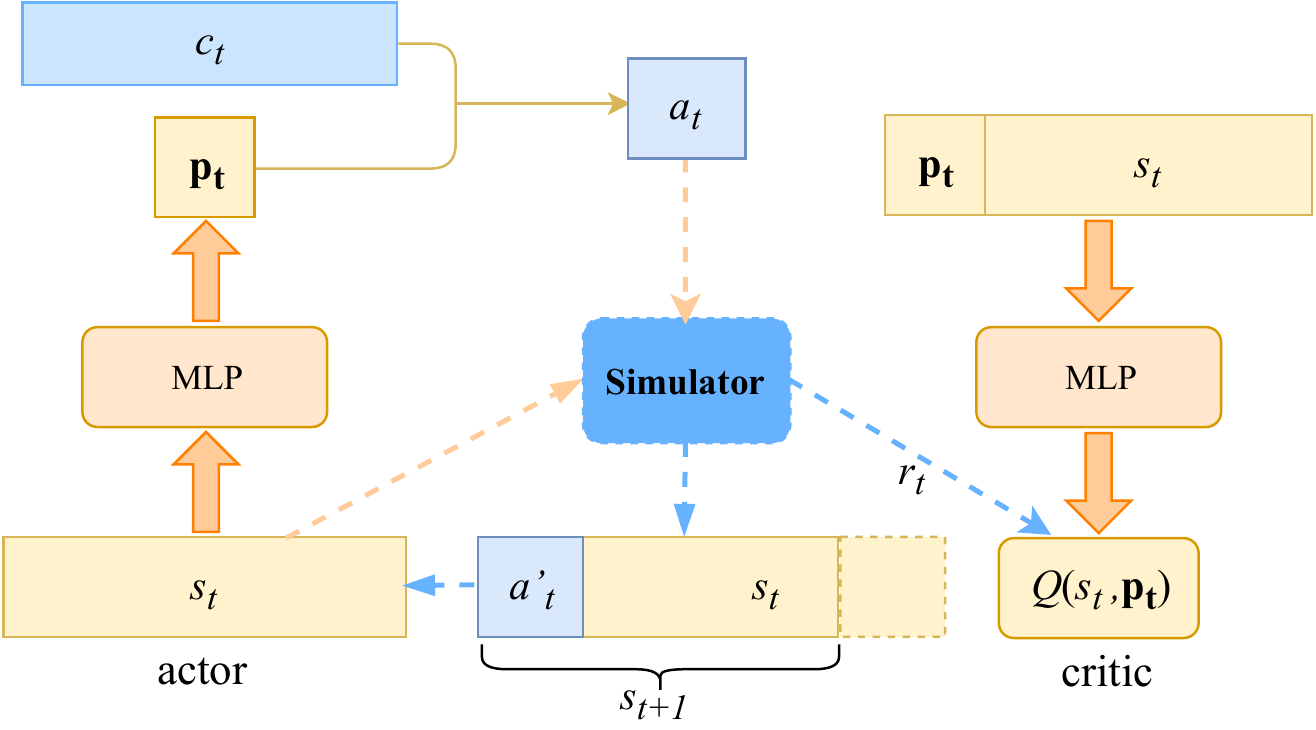}
	\caption{The Structure of TDDPG-Rec.}
	\label{fig: structure}
\end{figure}

\subsection{Environment Simulator}
It is expensive and time-consuming that utilizes real interactive environment for the training of RL model.
The same as several previous work~\cite{wei2017reinforcement,chen2019large}, based on history interactions, we build the environment simulator.
It receives the present state ${s}_{t}$ and action ${a}_{t}$, then returns reward ${r}_{t}$ and the next state ${s}_{t+1}$. In our model, the reward function guides the model to capture users' preferences and evaluate the rank quality of the recommended items. For user $u_i$ in time-step $t$, we give a reward ${r}_{t}$ on ${a}_{t}$ gained by $\textbf{\textrm{p}}_t$ among the candidate set $c_t$. The reward $r_t$ is determined by two values, ${w}_{k}$ and $\hat{y}_{i,j}$, specifically,
\begin{equation}
r_t = \mathcal{R}\left(s_t, {a}_{t}\right)=\sum\nolimits_{k=1}^{n_a} {w}_{k} \times \hat{y}_{i,j}
\end{equation}
where $n_a$ is the number of items in ${a}_{t}$, $w_{k}$ is the ranking weight of the items in $a_t$, and $\hat{y}_{i,j}$ is the adjusted rating of item $v_j$ for user $u_i$.

Inspired by DCG~\cite{Jarvelin2002,wei2017reinforcement}, the ranking weight is calculated by,
\begin{equation}
{w}_{k} = {1}/{\log _{2}(k+1)}
\end{equation}

To give proper rewards for different types of items, $\hat{y}_{i,j}$ is designed as follows,
\begin{equation}
\hat{y}_{i,j}=\left\{
\begin{array}{lll}
y_{i,j} - y_b, & \text{if}~~ v_j \in \mathcal{V}^p_{u_i}; \\
y_{i,j} - y_b -1 , &\text{if}~~ v_j \in \mathcal{V}^n_{u_i}; \\
-$0.5$, &\text{if}~~ v_j \in \mathcal{V}^p_{cl_l^f} - (\mathcal{V}^p_{cl_l^f} \cap (\mathcal{V}^p_{cl_{l}} \cup \mathcal{V}^n_{u_i})) ; \\
0, & \text{otherwise}.
\end{array}
\right.
\end{equation}
Recall here $y_{i,j}$ is the rating of user $u_i$ on item $v_j$, and $y_b$ is the rating bound to determine whether the corresponding log is regarded as positive or negative. By this formula, positive items in ${a}_{t}$ will get positive feedback, and negative items will get negative feedback. Moreover, the supplemented negative items will get half of the minimum negative feedback, i.e., $-0.5$, while the other items will get a feedback of $0$.

As shown in Figure~\ref{fig: structure}, our method of generating ${s}_{t+1}$ is in a sliding-window manner. Specifically, among ordered items in ${a}_{t}$, we keep the order and select the items that are not in ${s}_{t}$ as ${a}'_{t}$. Then we put ${a}'_{t}$ at the head of ${s}_{t}$, and select the top $n_s$ items as ${s}_{t+1}$.

\begin{algorithm}[t]
	\caption{Learning TDDPG-Rec}
	\KwIn{The maximum step $T$; The number of user clusters  $n_{cl}$; Target network update rate $\tau$.}
	\KwOut{Model parameters $\theta^{Q}$, $\theta^{\mu}$, $\theta^{Q^{\prime}}$ and $\theta^{\mu^{\prime}}$.}
	\For{$l=1$ to $n_{cl}$}{
		Randomly initialize critic network $Q\left(s, \textbf{\textrm{p}} | \theta^{Q}\right)$ and actor network $\mu\left(s | \theta^{\mu}\right)$\;
		Initialize target network: $\theta^{Q^{\prime}}_{l} = \theta^{Q}_{l}$, $\theta^{\mu^{\prime}}_{l} = \theta^{\mu}_{l}$\;
		Initialize replay buffer $\mathrm{B}$\;
		\Repeat{converge}{
			Randomly select $u_j$ in cluster $cl_l$\;
			Initialize observation state $s_1$\;
			Initialize a random process $\mathcal{N}$ for exploration\;
			\For {$t = 1$ to $T$}{
				Construct $c_t$ for $u_j$ by Algorithm 1\;
				$\textbf{\textrm{p}}_{t}=\mu\left(s_{t} | \theta^{\mu}\right)+\mathcal{N}_{t}$\;
				Use $\textbf{\textrm{p}}_{t}$ to select items from $c_t$ as $a_t$\;
				Interact with simulator by $a_t$\;
				Observe $r_t$ and $s_{t+1}$ from simulator\;
				Store transition $(s_t; \textbf{\textrm{p}}_t; r_t; s_{t+1})$ in $\mathrm{B}$\;
				Sample a random batch of $N_b$ transitions $(s_i; \textbf{\textrm{p}}_i; r_i; s_{i+1})$ from $\mathrm{B}$\;
				Update critic by minimizing Equation~(\ref{equ: loss})\;
				Update actor by Equation~(\ref{equ: pg})\;
				Update the target networks: $\theta^{Q^{\prime}} = \tau \theta^{Q}+(1-\tau) \theta^{Q^{\prime}}$, $\theta^{\mu^{\prime}} = \tau \theta^{\mu}+(1-\tau) \theta^{\mu^{\prime}}$\;
			}
		}
	}
	return $\theta^{Q}$, $\theta^{\mu}$, $\theta^{Q^{\prime}}$ and $\theta^{\mu^{\prime}}$
\end{algorithm}

\subsection{Learning TDDPG-Rec}
The training phase (as shown in Algorithm 2) learns model parameters $\theta^{Q}$, $\theta^{\mu}$, $\theta^{Q^{\prime}}$, and $\theta^{\mu^{\prime}}$ through maximizing the cumulated discounted rewards of all the decisions. Based on the assumption that similar users have similar preferences, our method classifies users and trains models for each cluster. At the beginning of the training phase, we randomly initialize the network parameters and the replay buffer $\mathrm{B}$. In order for action exploration, we initialize a random process $\mathcal{N}$, adding some uncertainty when generating $\textbf{\textrm{p}}$. The critic network focuses on minimizing the gap between the current Q-value $Q(s_{i}, \textbf{\textrm{p}}_i | \theta^{Q})$ and the expected Q-value $z_i$, which is measured by the following loss,
\begin{equation}
\label{equ: loss}
L(\theta^Q) = \frac{1}{N_b}  \sum\nolimits_{i}(z_i - Q(s_{i}, \textbf{\textrm{p}}_i | \theta^{Q}))^{2}
\end{equation}
where $z_i$ can be expressed in a recursive manner by using the Bellman equation,
\begin{equation}
z_i = r_{i}+\gamma Q^{\prime}(s_{i+1}, \mu^{\prime}(s_{i+1} | \theta^{\mu^{\prime}}) | \theta^{Q^{\prime}})
\end{equation}
The objective of the actor network is to optimize the policy vector $\textbf{\textrm{p}}$, through maximizing the Q-value. The actor network is trained by the sampled policy gradient:
\begin{equation}
\label{equ: pg}
\nabla_{\theta^{\mu}} J \approx \frac{1}{N_b} \sum\nolimits_{i} \nabla_{\textbf{\textrm{p}} } Q(s, \textbf{\textrm{p}}  | \theta^{Q})|_{s=s_{i}, \textbf{\textrm{p}} =\mu(s_{i})}
\nabla_{\theta^{\mu}} \mu(s | \theta^{\mu})|_{s_{i}}
\end{equation}

Note that in our implementation, we set the minimum and maximum training step thresholds based on the size of buffer $\mathrm{B}$. When the number of steps is greater than the minimum threshold and the loss remains stable, or the number of steps is greater than the maximum threshold, the training phase will stop.

\section{Experiments and Results}
\label{sec: experiments}

In this section, we conduct experiments to demonstrate the effectiveness of the proposed TDDPG-Rec model 
versus several state-of-the-art models. We first introduce the experimental setup, and then present and discuss the experimental results from the perspective of both recommendation performance and time efficiency. Finally, we conduct the hype-parameter sensitivity analysis in the last part of this section. We have implemented our  models based on  Tensorflow, which can be accessed in GitHub\footnote{https://github.com/SunwardTree/TDDPG-Rec}.

\subsection{Experimental Settings}
\subsubsection{Datasets}
Jure Leskovec \emph{et al.}~\cite{snapnets} collected and categorized a variety of {\it Amazon} products and built several datasets\footnote{http://snap.stanford.edu/data/amazon/productGraph/categoryFiles} including ratings, descriptions, and reviews. We evaluate our models on three publicly available {\it Amazon}  datasets: Digital Music (\textbf{Music} for short), \textbf{Beauty} and Clothing Shoes and Jewelry (\textbf{Clothing} for short), which all have at least $5$ reviews for each product. Table~\ref{datasets} shows the statistical details of the datasets we used.

\setlength{\tabcolsep}{2.4pt}
\begin{table}
	\scriptsize
	\centering
	\caption{Statistics of datasets.}
\vspace{-0.5em}
	\label{datasets}
	\begin{tabular}{lrrrrr}
		\hline
		\rule{0pt}{6pt}
		\textbf{DataSet} & \#\textbf{Users} & \#\textbf{Items} & \#\textbf{Ratings of Pos \& Neg} & \textbf{Sparsity} & \textbf{Size of Textual Info}\\
		\hline
		\rule{0pt}{6pt}
	Music & 5,541 & 3,568 & 58,905 ~~~5,801 & 0.9967 & 68,096 KB\\
Beauty & 22,363 & 12,101 & 176,520~~~21,982 & 0.9993 & 88,986 KB\\
		Clothing & 39,387 & 23,033 & 252,022~~~26,655 & 0.9997 & 84,168 KB\\
		\hline
	\end{tabular}
\end{table}

\begin{table*}
\begin{center}
	\tiny
	\caption{Overall interactive recommendation performance. Best performance is  in boldface and second best is underlined.}
	\vspace{-1em}
	\setlength{\tabcolsep}{1mm}{
		\resizebox{\textwidth}{15mm}{
			\begin{tabular}{c|cccccc|cccccc|cccccc}
				\hline
				\rule{0pt}{6pt}
				\multirow{2}{*}{\textbf{Method}} & \multicolumn{6}{c|}{\textbf{Music}}           & \multicolumn{6}{c|}{\textbf{Beauty}}          & \multicolumn{6}{c}{\textbf{Clothing}} \\
				\cline{2-19}
				\rule{0pt}{6pt}
				& {HR@10} & {F1@10} & \multicolumn{1}{l}{{nDCG@10}} & {HR@20} & {F1@20} & {nDCG@20} & {HR@10} & {F1@10} & {nDCG@10} & {HR@20} & {F1@20} & {nDCG@20} & {HR@10} & {F1@10} & {nDCG@10} & {HR@20} & {F1@20} & {nDCG@20} \\
				\hline
				\rule{0pt}{6pt}
				{ItemPop} & 0.2447  & 0.0454  & 0.1101  & 0.4889  & 0.0525  & 0.1716  & 0.2551  & 0.0482  & 0.1134  & 0.5278  & 0.0543  & 0.1817  & 0.2265  & 0.0413  & 0.1033  & 0.4964  & 0.0482  & 0.1706  \\
				{LinerUCB} & 0.3318  & 0.0621  & 0.1569  & 0.5885  & 0.0626  & 0.2210  & 0.2734   & 0.0502  & 0.1249  & 0.5273  & 0.0529 & 0.1885 & 0.2393  & 0.0437  & 0.1041  & 0.5044  & 0.0488  & 0.1704  \\
				{DMF} & 0.3201  & 0.0631  & 0.1462  & 0.5747  & 0.0638  & 0.2095  & 0.3219  & 0.0614  & 0.1447  & 0.5911  & 0.0613  & 0.2122  & 0.2500  & 0.0458  & 0.1130  & 0.5041  & 0.0489  & 0.1756  \\
				{ANR} & 0.4980  & 0.1128  & 0.2756  & 0.7097  & 0.1084  & 0.3252  & 0.4550  & 0.0990  & 0.2252  & 0.6993  & 0.1006  & 0.2850  & 0.3421  & 0.0663  & 0.1622  & 0.6008  & 0.0659  & 0.2264  \\
				{Caser} & 0.8097  & 0.1676  & 0.5351  & 0.9090  & 0.1048  & 0.5542  & 0.6125  & 0.1218  & 0.3939  & 0.7817  & 0.0826  & 0.4344  & 0.5060  & 0.0934  & 0.2900  & 0.7196  & 0.0702  & 0.3427  \\
				{SASRec} & \underline{0.8897} & \underline{0.1910}  & \underline{0.6212} & \textbf{0.9635}  & \textbf{0.1151}  & \underline{0.6325}  & \underline{0.6823}  & \underline{0.1386}  & \underline{0.4569}  & \underline{0.8330}  & \underline{0.0907}  & \underline{0.4942}  & \underline{0.5817}  & \underline{0.1084}  & \underline{0.3525}  & \underline{0.7655}  & \underline{0.0758}  & \underline{0.3968}  \\
				{DDPG-$k$NN~($k$=0.1\emph{M})} & 0.3274  & 0.0648  & 0.1527  & 0.5838  & 0.0647  & 0.2171  & 0.2585  & 0.0489  & 0.1170  & 0.5142  & 0.0529  & 0.1809  & 0.2541  & 0.0467  & 0.1131  & 0.5043  & 0.0490  & 0.1757  \\
				{DDPG-$k$NN~($k$=\emph{M})} & 0.3436  & 0.0692  & 0.1617  & 0.6001  & 0.0676  & 0.2258  & 0.2734  & 0.0522  & 0.1274  & 0.5197  & 0.0539  & 0.1889  & 0.2768  & 0.0510  & 0.1242  & 0.5293  & 0.0517  & 0.1874  \\
				\hline
				\rule{0pt}{6pt}
				{TDQN-Rec} & 0.8150  & 0.1712  & 0.5048  & 0.8977  & 0.1053  & 0.5251  & 0.5466  & 0.1053  & 0.3033  & 0.7662  & 0.0796  & 0.3573  & 0.3739  & 0.0685  & 0.1873  & 0.6252  & 0.0607  & 0.2501  \\
				{MDDPG-Rec} & 0.8293  & 0.1803 & 0.5293  & 0.9074  & 0.1087 & 0.5477  & 0.5985  & 0.1202  & 0.3391  & 0.7740  & 0.0825 & 0.3830  & 0.3094  & 0.0567  & 0.1487  & 0.5595  & 0.0542  & 0.2111  \\
				{TDDPG-Rec} & \textbf{0.9164}  & \textbf{0.2032}  & \textbf{0.6630}  & \underline{0.9426} & \underline{0.1142}  & \textbf{0.6687} & \textbf{0.7648} & \textbf{0.1517} & \textbf{0.4942} & \textbf{0.8952} & \textbf{0.0948}  & \textbf{0.5261} & \textbf{0.6237} & \textbf{0.1203} & \textbf{0.3553} & \textbf{0.8210 } & \textbf{0.0818} & \textbf{0.4055} \\
				\hline
		\end{tabular}}
	}
	\label{tab:results}
\end{center}
\end{table*}

\subsubsection{Baseline methods}
We compare TDDPG-Rec with $10$ methods, where ItemPop is a conventional recommendation method, LinearUCB is a MAB-based method, DMF is an MF-based method with neural network, ANR is a neural recommendation method that leverages textual information, Caser and SASRec are time-related deep learning based methods, DDPG-KNN, TPGR, TDQN-Rec and MDDPG-Rec are all RL-based methods.
\begin{itemize}
	\item \textbf{ItemPop}~recommends the most popular items (i.e., the item with the highest average rating) from currently available items to the user at each timestep. This method is non-personalized and is often used as a benchmark for recommendation tasks.
	
	\item  \textbf{LinearUCB}~\cite{li2010contextual} is a contextual-bandit recommendation approach that adopts a linear model to estimate the upper confidence bound for each arm.
	
	\item \textbf{DMF}~\cite{xue2017deep} is a state-of-the-art matrix factorization model using deep neural networks. Specifically, it utilizes two distinct MLPs to map the users and items into a common low-dimensional space with non-linear projections. 
	
	\item \textbf{ANR}~\cite{chin2018anr} uses an attention mechanism to focus on the relevant parts of reviews and estimates aspect-level user and item importance in a joint manner.
	
	\item \textbf{Caser}~\cite{Tang2018} embeds a sequence of recent items into an image and learn sequential patterns as local features of the image by using convolutional filters.
	
	\item \textbf{SASRec}~\cite{McAuley2018}  is a self-attention based sequential model for next item recommendation. It models the entire user sequence and adaptively considers consumed items for prediction.
	
	\item \textbf{DDPG-$k$NN}~\cite{dulac2015deep} addresses the large discrete action space problem by combining DDPG with an approximate $k$NN method.
	
	\item \textbf{TPGR}~\cite{chen2019large} builds a balanced hierarchical clustering tree and formulates picking an item as seeking a path from the root to a certain leaf of the tree.
	
	\item \textbf{TDQN-Rec} is a method that replaces DDPG in TDDPG-Rec with DQN, while retains other components  the same as that in TDDPG-Rec.
	
	\item \textbf{MDDPG-Rec} is a method that uses the same framework as TDDPG-Rec, but with vectors being derived by matrix factorization~\cite{koren2009matrix}, rather than leveraging textual information.
\end{itemize}

Note that for DDPG-$k$NN, larger $k$ (i.e., the number of nearest neighbors) will result in better performance but poor efficiency. For a fair comparison, we consider setting $k$ as $0.1M$ and $M$ (recall that $M$ is the number of items) respectively.  

\subsubsection{Evaluation Metrics and Methodology}
The methods that achieve their goals by Top-$k$ recommendation take evaluation on the indexes like Hit Ratio (HR)~\cite{xue2017deep}, Precision~\cite{zheng2018drn,Tang2018}, Recall~\cite{Tang2018},  F1~\cite{chen2019large}  and normalized Discounted Cumulative Gain (nDCG)~\cite{wei2017reinforcement,zhao2018deep,zheng2018drn,McAuley2018}. To cover as many aspects of Top-$k$ recommendation as possible, we chose HR@$k$, F1@$k$, and nDCG@$k$ as the evaluation metrics.

The test data was constructed in data preparation, and all the evaluated methods were tested by using this data. We now describe the test method in detail: For each user, we first classify user's history logs into positive and negative ones, and sort the items in positive history logs by time-stamp. Then, we choose the last $10\%$ of the ordered items in the positive logs as positive items. Finally, the negative items are randomly selected from the cluster that is farthest from the one that the current user belongs to. Based on such a strategy, the recommendation methods (except TPGR, which only recommends one item in each episode) can generate a ranked Top-$k$ list to evaluate the metrics mentioned above.

\subsection{Results and Analysis}
Table~\ref{tab:results} shows the summarized results of our experiments on the three datasets in terms of six metrics including HR@$10$, F1@$10$, nDCG@$10$, HR@$20$, F1@$20$ and nDCG@$20$. Note that since TPGR  is not suitable for Top-$k$ recommendation,  we did not include it as a competitor when evaluating the recommendation performance. From the results, we have the following key observations:

\begin{itemize}
	\item The proposed model TDDPG-Rec achieves the best (or the second-best, but with a small gap to the best one) performance and obtains remarkable improvements over the state-of-the-art methods. Moreover, the performance improvement increases along with the increase of data scale and data sparsity, where the datasets are arranged in increasing order of scale and sparsity. This justifies the effectiveness of TDDPG-Rec that leverages textual information in RL-based recommendation, especially for large-scale and high-sparsity datasets.
	
	\item The structure of ANR is similar to that of DMF, while the structure of TDDPG-Rec is the same as that of MDDPG-Rec. Text-based methods ANR and TDDPG-Rec  consistently outperform their counterparts, DMF and MDDPG-Rec,  which only use interaction information for embedding. This demonstrates the importance of utilizing textual information to alleviate the negative effects of data sparsity for better performance.
\end{itemize}

Moreover, based on the results in Table~\ref{tab:results}, we conduct several statistical significance tests~\cite{Kulinskaya2011}, with significance level $\alpha = 0.05$. For all metrics, the $p$-value of SASRec and TDDPG-Rec is $6.9e^{-4} < 0.01$, the $p$-value of TDQN and TDDPG-Rec is $2.17e^{-5} < 0.01$, and the $p$-value of MDDPG-Rec and TDDPG-Rec is $3.69e^{-5} < 0.01$. The results indicate that there are significant differences between the evaluated couple methods.

\begin{table}
	\footnotesize
	\centering
	\caption{Time Comparison for training and decision making.}
\vspace{-1em}
	\label{tc}
	\begin{center}
		\begin{tabular}{c|c|c}
			\hline
			\rule{0pt}{6pt}
			{Method}  
			& \multirow{1}{*}{Per training step (ms)} & \multirow{1}{*}{Per decision (ms)} \\
			\hline
			\rule{0pt}{6pt}
			DDPG-$k$NN ($k=$0.1\emph{M}) & 5.26  & 7.86  \\
			DDPG-$k$NN ($k=$\emph{M}) & 5.51  & 56.42  \\
			TPGR  & 4.98  & 5.06  \\
			SASRec & 205.12  & 14.32  \\
			TDQN-Rec  & 1.86  & 1.08  \\
			TDDPG-Rec & 3.13  & 0.90  \\
			\hline
		\end{tabular}
	\end{center}
	\label{tab:tc}
\end{table}

\subsection{Time Comparison}
In this section, we compare the efficiency of RL-based models on \textbf{Beauty} dataset from two aspects, the consumed time of training (updating the model) and decision making (selecting action), where the time spend is measured in millisecond. Note that since SASRec provides competitive results, we also include SASRec as a competitor. To make a fair comparison, both $n_s$ and $n_a$ are set to $1$, and the experiments are conducted on the same machine with 6-core CPU (i7-6850k, 3.6GHz) and 64GB RAM.

As shown in Table~\ref{tab:tc}, DDPG-$k$NN  runs much slower than other models, because it has high time complexity. TDQN-Rec consumes the least time on training, due to its simple structure. But as shown in Table~\ref{tab:results}, it has the worst recommendation performance among all the RL-based methods. TPGR reduces the decision-making time significantly by constructing a clustering tree, but as mentioned before, it only supports Top-$1$ recommendation. Compared to other methods, by using policy vector and action candidate, our model TDDPG-Rec achieves significant improvement in terms of both recommendation performance and efficiency.

\subsection{Parameter Sensitivity}
We select several important parameters to analyze their effects on the performance of TDDPG-Rec in terms of HR@$10$ and nDCG@$10$. Note that we have conducted such experiments on all the datasets, and the results show that our approach exhibits similar performance trends on all the evaluated datasets. For simplicity, we only present the results in \textbf{Beauty} dataset. When testing one parameter, we keep the others fixed. The default settings are $D=100$, $n_{cl}=5$, $n_c=100$, $\alpha=0.5$, $n_s=20$ and $n_a=5$.

\begin{figure}[t]
	\centering
	\subfloat[\scriptsize{}]{
		\begin{minipage}[t]{0.23\textwidth}
			\centering
			\includegraphics[width=\linewidth, height=0.65\linewidth]{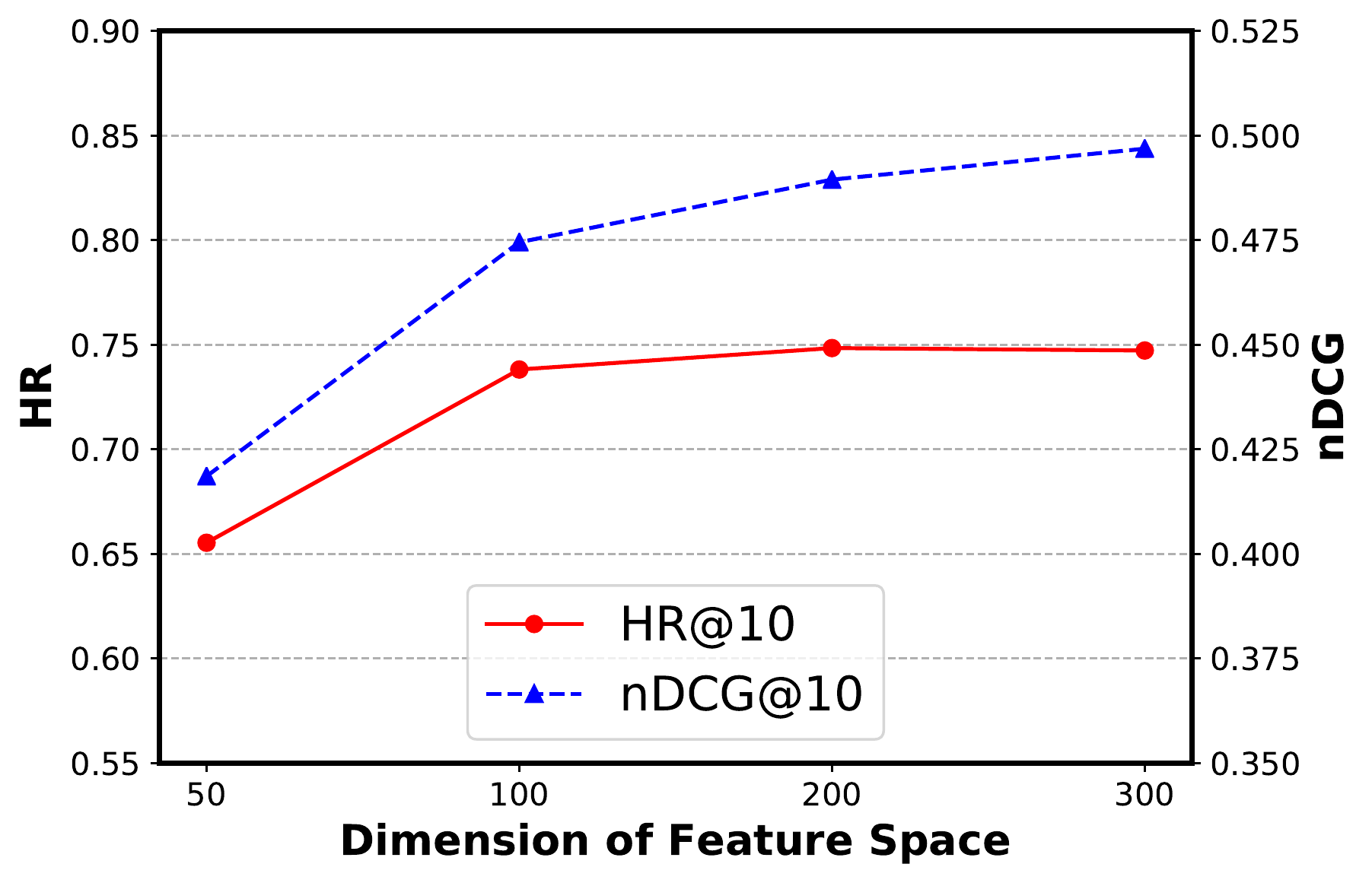}
		\end{minipage}
	}
	\subfloat[\scriptsize{}]{
		\begin{minipage}[t]{0.23\textwidth}
			\centering
			\includegraphics[width=\linewidth, height=0.65\linewidth]{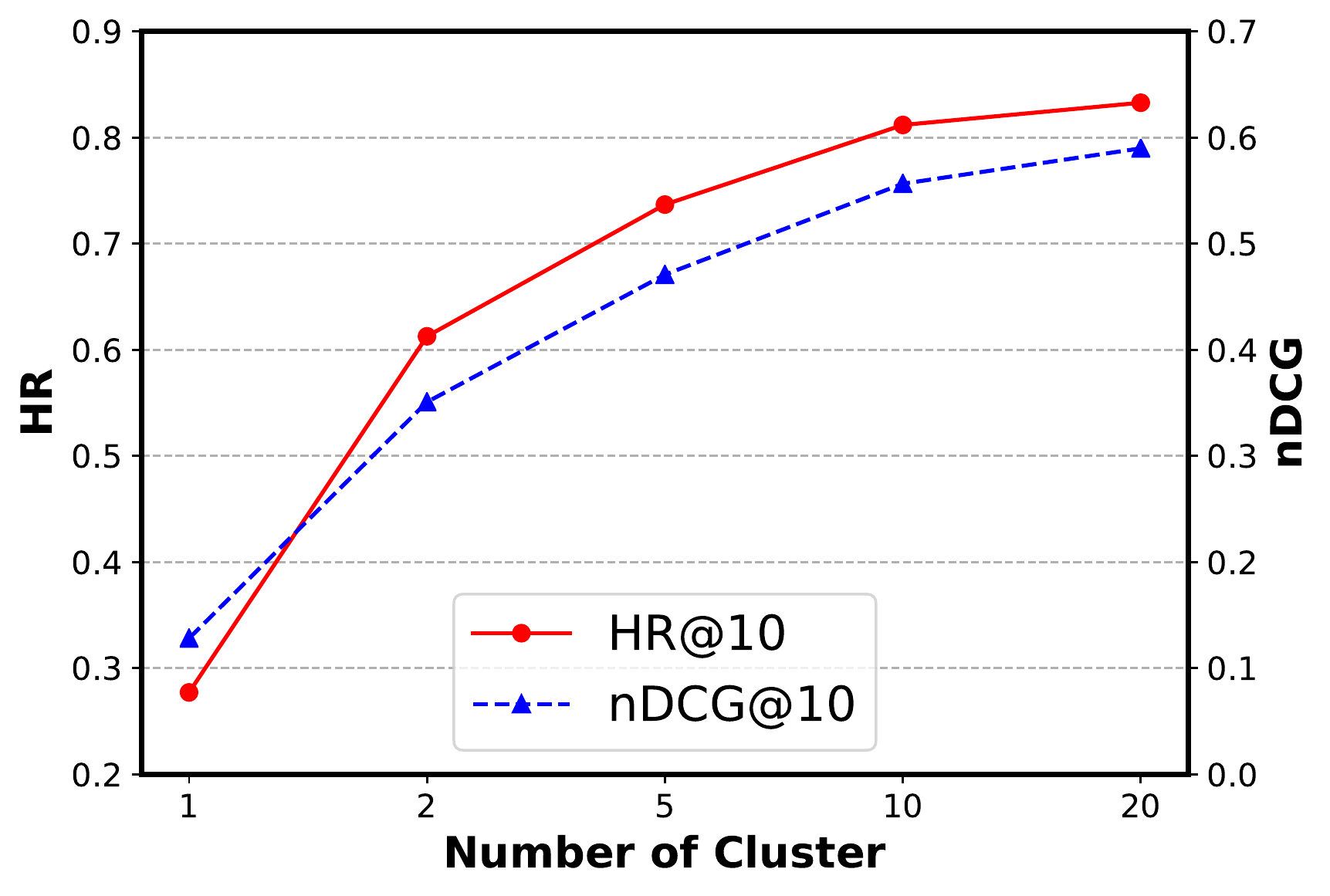}
		\end{minipage}
	}\\  
	\caption{Performance of HR@10 and nDCG@10 \emph{w.r.t.}  (a) the dimension of feature space; (b) the number of clusters.}
	\label{fig: p1}
\end{figure}

\vspace{0.5em}
\noindent \textbf{The Dimension of Feature Space ($D$)}
~~The number of feature dimension reflects the richness of the information. As shown in~Figure~\ref{fig: p1} (a), with the increase of $D$, as expected, TDDPG-Rec also performs better.

\vspace{0.5em}
\noindent\textbf{The Number of Clusters ($n_{cl}$)}
~~As shown in Figure~\ref{fig: p1}\;(b), the increase of $n_{cl}$ improves the performance. This is mainly because the more clusters, the larger the difference between the current cluster and the one farthest from it, which leads to more high-quality negative items, and eventually results in better performance.

\begin{figure}[t]
	\centering
	\subfloat[\scriptsize{}]{
		\begin{minipage}[t]{0.23\textwidth}
			\centering
			\includegraphics[width=\linewidth, height=0.65\linewidth]{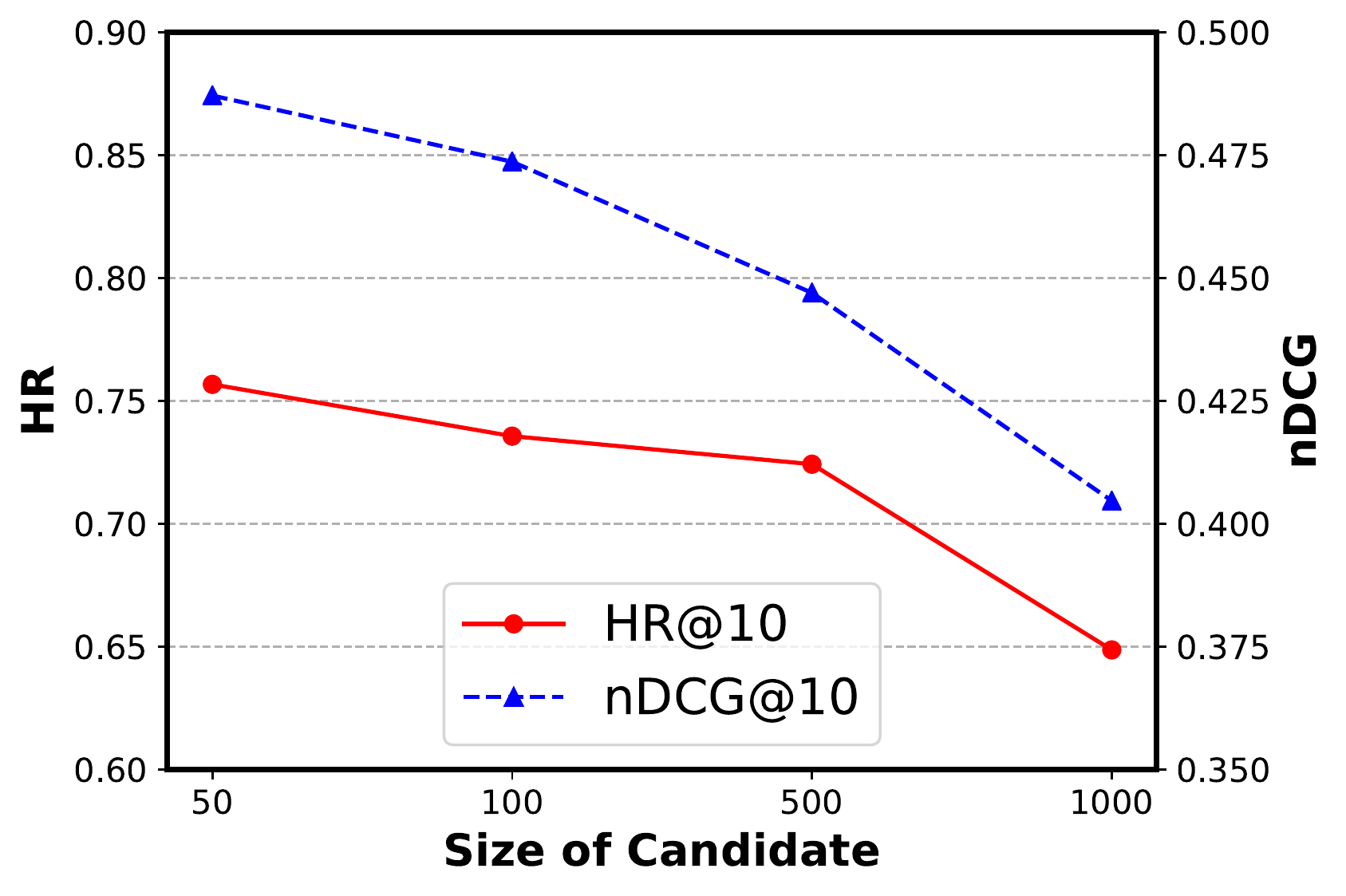}
		\end{minipage}
	}
	\subfloat[\scriptsize{}]{
		\begin{minipage}[t]{0.23\textwidth}
			\centering
			\includegraphics[width=\linewidth, height=0.65\linewidth]{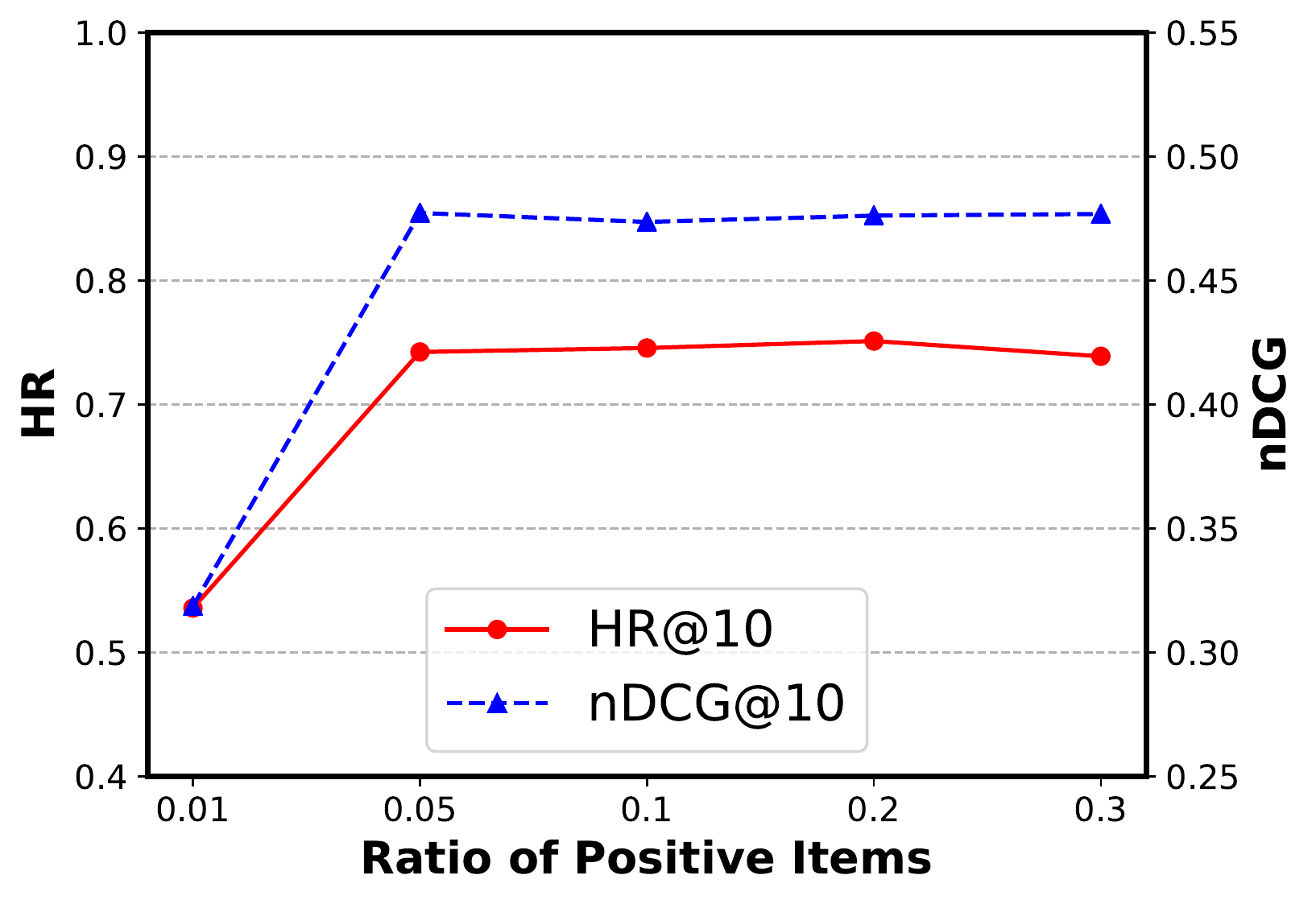}
		\end{minipage}
	}\\	
	\caption{Performance of HR@10 and nDCG@10 \emph{w.r.t.}  (a) the size of candidate; (b) the ratio of positive items.}
	\label{fig: p2}
\end{figure}

\vspace{0.5em}
\noindent\textbf{The Size of Candidate ($n_c$)}
~~Figure~\ref{fig: p2}\;(a) shows that the performance decreases with the increase of $n_c$. This is mainly because the items in $\mathcal{V}^p_{u_i}$ are much less than the items in $\mathcal{V}^p_{cl_l^f}$. In other words, the increase of $n_c$ will cause imbalance sampling, which in turn leads to worse performance.

\vspace{0.5em}
\noindent\textbf{The Ratio of Positive Items ($\alpha$)}
~~As shown in Figure~\ref{fig: p2}\;(b), with the increase of $\alpha$, the performance first grows and then remains stable. This is because increasing $\alpha$ will introduce more positive items to perceive the user's interests better. But since $n_{pos} \leq |\mathcal{V}^p_{u_i}|$ (see Algorithm 1), when $\alpha$ is big enough, its growth may no longer affect $n_{pos}$.

\begin{figure}[t]
	\centering
	\subfloat[\scriptsize{}]{
		\begin{minipage}[t]{0.23\textwidth}
			\centering
			\includegraphics[width=\linewidth, height=0.65\linewidth]{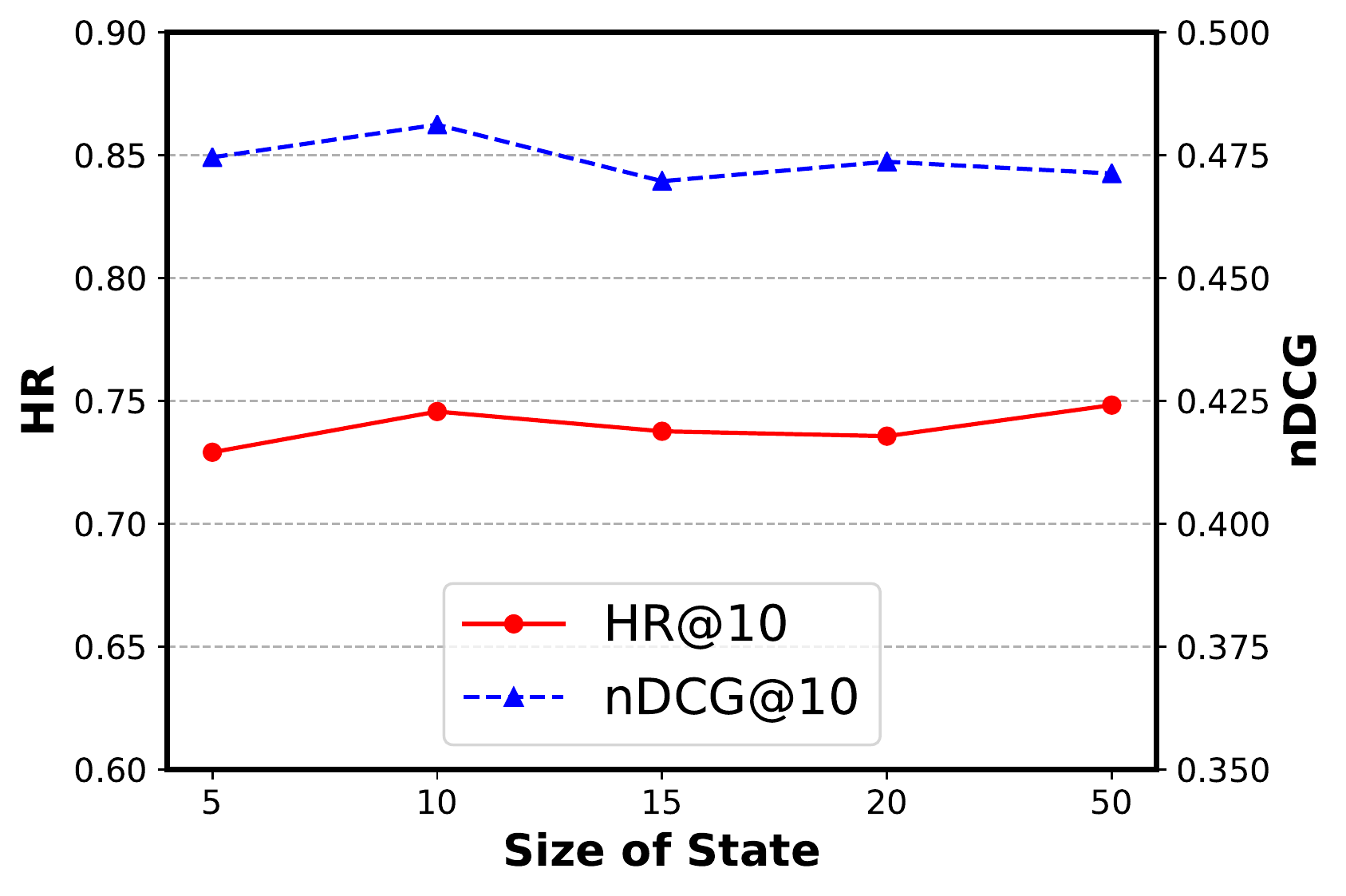}
		\end{minipage}
	}
	\subfloat[\scriptsize{}]{
		\begin{minipage}[t]{0.23\textwidth}
			\centering
			\includegraphics[width=\linewidth, height=0.65\linewidth]{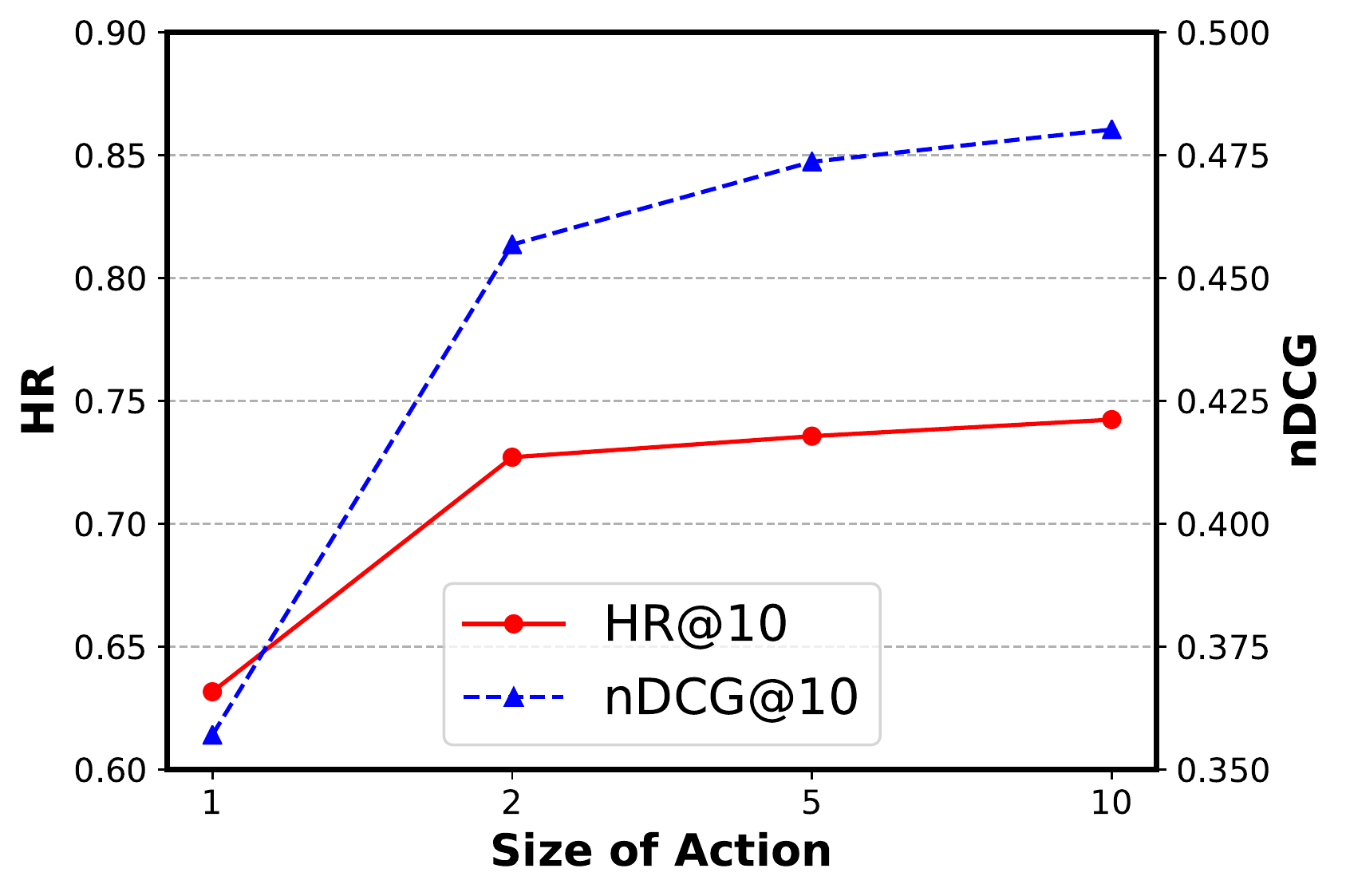}
		\end{minipage}
	}\\  
	\caption{Performance of HR@10 and nDCG@10 \emph{w.r.t.}  (a) the size of state; (b) the size of action.}
	\label{fig: p3}
\end{figure}

\vspace{0.5em}
\noindent\textbf{The Size of State ($n_s$)}
~~Figure~\ref{fig: p3}\;(a) shows that the performance stays smoothly with the increase of $n_s$, which means the size of state impacts TDDPG-Rec little.

\vspace{0.5em}
\noindent\textbf{The Size of Action ($n_a$)}
~~Figure~\ref{fig: p3}\;(b) shows that with the increase of $n_a$, the performance also increases. This is due to that the larger $n_a$ is, the more frequent the user state changes, which makes the positive items have more opportunities to be selected.

\section{Conclusion}
\label{sec: conclusion}
In this paper, we propose TDDPG-Rec, a Text-based Deep Deterministic Policy Gradient framework for Top-$k$ interactive recommendation. By leveraging textual information and pre-trained word vectors, we embed items and users into a same feature space,  which greatly alleviates the data sparsity problem. Moreover, based on the thought of collaborative filtering, we classify users into several clusters and construct an action candidate set. Combining with the policy vector dynamically learned from DDPG that expresses the user's preferences in the feature space, we select items from the candidate set to form action for recommendation, which greatly improves the efficiency of decision making.  Experimental results over a carefully designed simulator on three public datasets demonstrate that compared with state-of-the-art methods, TDDPG-Rec can achieve remarkable performance improvement in a time-efficient manner. 

For future work, we would like to see whether utilizing other techniques, such as the attention mechanism, can achieve better recommendation accuracy. Moreover, we intend to study if it is possible to incorporate our proposed model with transfer learning.

\ack We would like to thank the referees for their valuable comments, which
helped improve this paper considerably. The work was partially supported by the National Natural Science Foundation of China under Grant No. 61672252, and the Fundamental Research Funds for the Central Universities under Grant No. 2019kfyXKJC021.

\bibliography{ecai}

\end{document}